\documentstyle[12pt]{article}

\title{\textbf{the Relative Entropy as an Increasing Function of Time in Cosmology}}
\author{\textbf{Cheng-Yi Sun\footnote{cysun@mails.gucas.ac.cn}
and De-Hai Zhang\footnote{dhzhang@gucas.ac.cn}}\\
Department of Physics,\\
The Graduate School of The Chinese Academy of Sciences,\\
Beijing 10049, P.R.China.}

\begin{document}
\maketitle
\begin{abstract}
In this paper, it is shown that the relative entropy is an
increasing function of time in both the linear regime and the
non-linear regime during the large scale structure formation in
cosmology.
\end{abstract}



\section{Introduction}
In \cite{g0402076} the authors have proposed a measure which
quantifies the distinguishability of the actual mass distribution
from its spatial average, borrowing a well-known concept in
standard information, the \emph{relative entropy},
\begin{equation}
\label{primaryDef}S\{p||q\}=\sum_{i}p_i\ln\frac{p_i}{q_i},
\end{equation}
where $\{q_i\}$ is the probability distribution and $\{p_i\}$ is
the actual one. For a continuum the relevant quantity is
\begin{equation}
\label{defForContiuum}\frac{S\{\rho||\langle\rho\rangle_D\}}{V_D}=\langle
\rho\ln\frac{\rho}{\langle\rho\rangle_D}\rangle_D,
\end{equation}
where $\rho$ is the actual distribution and
$\langle\cdots\rangle_D$ is its spatial average in the volume
$V_D$ on the compact domain $D$ of the manifold $\Sigma$. For
scalar functions $\Psi(t,X^i)$, the averaging operation, in term
of Riemannian volume integration, is defined as
\begin{equation}
\label{averageOpe}\langle\Psi(t,X^i)\rangle_D:=\frac{1}{V_D}\int_D{\sqrt{g}d^3X\
\Psi(t,X^i)},
\end{equation}
with $g:=det(g_{ij})$ and the volume of an arbitrary compact
domain, $V_D(t):=\int_D{\sqrt{g}d^3X}$; $X^i$ are coordinates in a
$t=const.$ hypersurface (with 3-metric $g_{ij}$) that are comoving
with fluid elements of dust:
\begin{equation}
\label{metric}ds^2=-dt^2+g_{ij}dX^idX^j.
\end{equation}
The derivative of the relative entropy with respect to the time is
given in \cite{g0402076} by
\begin{equation}
\label{derivativeOfRE}\frac{\dot{S}\{\rho||\langle\rho\rangle_D\}}{V_D}
=-\langle\delta\rho\theta\rangle_D,
\end{equation}
where $\theta$ denotes the local expansion rate (as the trace of
the extrinsic curvature of the hypersurfaces $t=const.$).
Hereafter, a ``dot" denotes the derivative respect the physical
time, $\dot{\Psi}=d\Psi/dt$ and $\delta$ denotes the derivative of
a local field from its spatial average,
$\delta\Psi=\Psi-\langle\Psi\rangle_D$.

In \cite{g0402076}, a conjecture is given: \emph{The Relative
Information Entropy of a dust matter model
$S\{\rho||\langle\rho\rangle_D\}$ is, for sufficiently large
times, globally (i.e. averaged over the whole compact manifold
$\Sigma$) an increasing function of time}.

In the cosmology, the `sufficiently large times' means after the
beginning of the non-linear regime during the structure formation.
In this regime, the overdense elements ($\delta\rho>0$) stop
expanding ($\theta\leq 0$) and the underdense elements
($\delta\rho<0$) expand ($\theta>0$). So we see the conjecture is
plausible.

In this paper it will be shown that, in cosmology, the relative
entropy is globally an increasing function of time even in the
linear regime during the structure formation. It should be noted
that this point can not be concluded from
Eq.(\ref{derivativeOfRE}), because during the linear regime, the
overdense elements ($\delta\rho>0$) are also expanding
($\theta>0$).

Below, firstly an proof of the positivity of the entropy,
$S\{\rho||\langle\rho\rangle_D\}$ is shown. Then it is shown that
the entropy is globally an increasing function of time even in the
linear regime during the structure formation.

\section{the Positivity of the Relativity Entropy}
We first express the relative entropy in term of the integral,
\begin{equation}
\label{integralFormOfRE}S\{\rho||\langle\rho\rangle_D\}=\int_D{\sqrt{g}d^3X\
\rho\ln\frac{\rho}{\langle\rho\rangle_D}}.
\end{equation}
For an overdense volume element, its density may be taken as
$\rho=\langle\rho\rangle_D+\delta\rho, \delta\rho>0$. In order to
ensure the average density to be $\langle\rho\rangle_D$, we may
divide the overdense elements into three types. For an element of
the first type with $\rho=\langle\rho\rangle_D+\delta\rho$, there
must exist one or several underdense elements that their densities
can be expressed as $\rho=\langle\rho\rangle_D-h_i\delta\rho$,
with $h_i>0, \sum_{i}h_i=1, i=1,2,\cdots$. The region where the
elements of the first type are localized is denoted by
$\tilde{D}_1$.

For an overdense element of the second type, such one or several
underdense elements does not exist. However, corresponding to
several elements of the second type, there must exist one
underdense element. Writing the density of this underdense element
as $\rho=\langle\rho\rangle_D-\delta\rho$, the densities of the
several overdense elements can be expressed as
$\rho=\langle\rho\rangle_D+g_i\delta\rho$, with $g_i>0,
\sum_{i}{g_i}=1, i=2,3,\cdots$. The region where these underdense
elements are localized is denoted by $\tilde{D}_2$. In fact, here
we have given an correspondence between the overdense elements and
the underdense elements by giving $h_i$ and $g_i$.

Then Eq.(\ref{integralFormOfRE}) can be written as
\begin{eqnarray}
S\{\rho||\langle\rho\rangle_D\}&=&\int_{\tilde{D}_1}{\sqrt{g}d^3X\
f_1(\langle\rho(t,X^i)\rangle_D,\delta\rho(t,X^i))} \nonumber\\
&+&\int_{\tilde{D}_2}{\sqrt{g}d^3X\
f_2(\langle\rho(t,X^i)\rangle_D,\delta\rho(t,X^i))},\label{overdenseRE}\\
f_1(\rho_0,\delta\rho)
&=&(\rho_0+\delta\rho)\ln\frac{\rho_0+\delta\rho}
{\rho_0}+\sum_{i}(\rho_0-h_i\delta\rho)
\ln\frac{\rho_0-h_i\delta\rho}{\rho_0}, \label{integralFun1}\\
f_2(\rho_0,\delta\rho)
&=&\sum_{i}(\rho_0+g_i\delta\rho)\ln\frac{\rho_0+g_i\delta\rho}
{\rho_0}+(\rho_0-\delta\rho)\ln\frac{\rho_0-\delta\rho}{\rho_0},
\label{integralFun2}
\end{eqnarray}
where $\rho_0\equiv\langle\rho\rangle_D$. The derivatives of the
two functionals, $f_1(\langle\rho\rangle_D,\delta\rho)$ and
$f_2(\langle\rho\rangle_D,\delta\rho)$, with respect to
$\delta\rho$, are
\begin{eqnarray}
\frac{\delta
f_1}{\delta(\delta\rho)}&=&\ln\frac{\langle\rho\rangle_D+\delta\rho}
{\langle\rho\rangle_D}+\sum_{i}(-f_i)\ln\frac{\langle\rho\rangle_D-f_i\delta\rho}
{\langle\rho\rangle_D}>0, \label{derivativeOfF1}\\
\frac{\delta
f_2}{\delta(\delta\rho)}&=&\sum_{i}g_i\ln\frac{\langle\rho\rangle_D+g_i\delta\rho}
{\langle\rho\rangle_D}-\ln\frac{\langle\rho\rangle_D-\delta\rho}
{\langle\rho\rangle_D}>0. \label{derivativeOfF2}
\end{eqnarray}
We have used the conditions $\sum_{i}h_i=1$ and $\sum_{i}g_i=1$.
Then two inequalities are obtained:
$f_1(\langle\rho\rangle_D,\delta\rho)\geq
f_1(\langle\rho\rangle_D,0)=0$ and
$f_2(\langle\rho\rangle_D,\delta\rho)\geq
f_2(\langle\rho\rangle_D,0)=0$. Now, due to
Eq.(\ref{overdenseRE}), it can be concluded that the relative
entropy, $S\{\rho||\langle\rho\rangle_D\}$, is positive.

\section{the Relative Entropy as an Increasing Function of Time in Linear Regime}

In linear regime, both $\delta\rho$ and $\langle\rho\rangle_D$ are
decreasing functions of time. Then, due to the results in the last
section, it cannot be concluded whether
$S\{\rho||\langle\rho\rangle_D\}$ is an increasing function or
not. However, it is well known that the density contrast on the
overdense region:
$\delta\equiv\frac{\delta\rho}{\langle\rho\rangle_D}$, is an
increasing function of time in linear regime. Although the density
contrast on the underdense region is an decreasing function of
time in linear regime, let's express the density contrast on the
underdense region as
$-\delta=-\frac{\delta\rho}{\langle\rho\rangle_D}$ with
$\delta\rho>0$ and $\delta>0$. Then $\delta$ on the underdense
region is also an increasing function of time in linear regime.
Now, rewrite the relative entropy in terms of the density contrast
as
\begin{eqnarray}
S\{\rho||\langle\rho\rangle_D\}&=&\int_{\tilde{D}_1}{d^3X\
\sqrt{g}\langle\rho(t,X^i)\rangle_D\tilde{f}_1(\delta(t,X^i))} \nonumber\\
&+&\int_{\tilde{D}_2}{d^3X\
\sqrt{g}\langle\rho(t,X^i)\rangle_D\tilde{f}_2(\delta(t,X^i))},\label{densityConRE}\\
\tilde{f}_1(\delta)
&\equiv&(1+\delta)\ln(1+\delta)+\sum_{i}(1-h_i\delta)\ln(1-h_i\delta).
\label{densityConFun1}\\
\tilde{f}_2(\delta)
&\equiv&\sum_{i}(1+g_i\delta)\ln(1+g_i\delta)+(1-\delta)\ln(1-\delta).
\label{densityConFun2}
\end{eqnarray}
The derivatives of the two functionals, $\tilde{f}_1(\delta)$ and
$\tilde{f}_2(\delta)$, with respect to $\delta$, are
\begin{eqnarray}
\tilde{f}_1'(\delta)&=&\ln(1+\delta)+\sum_{i}(-h_i)\ln(1-h_i\delta)>0,
\label{derivativeOfFT1}\\
\tilde{f}_2'(\delta)&=&\sum_{i}g_i\ln(1+g_i\delta)+\ln(1-\delta)>0,
\label{derivativeOfFT2}
\end{eqnarray}
We have used the conditions $\sum_{i}h_i=1$ and $\sum_{i}g_i=1$.
Additional, we have also used the equations $\frac{\delta
h_i}{\delta(\delta)}=\frac{\delta g_i}{\delta(\delta)}=0$. This
implies we assumes that $h_i$ and $g_i$ are independent of
$\delta$. We believe this is reasonable. The reason is that the
changes in $h_i$ or $g_i$ just give another different
correspondence between the overdense elements and the underdense
elements. However, this does not change the relative entropy if
the distribution of the density is unchanged. Now we know
$\tilde{f}_1(\delta)$ and $\tilde{f}_2(\delta)$ are increasing
functionals of $\delta$. At the same time, for a dust matter
model, the variable $\sqrt{g}\langle\rho(t,X^i)\rangle_D$ is an
constant of time (This point has been indicated in
\cite{g0402076}.). Now, according to $\delta$ as an increasing
function of time  and Eq.(\ref{densityConRE}), it can be concluded
that, in cosmology, the Relative Entropy is globally an increasing
function of time in linear regime during the structure formation.

\section{Conjecture and Discussion}

Above, the relative entropy as an increasing function of time in
linear regime in cosmology has been shown. The process is not same
as in \cite{g0402076}. The derivative of the relative entropy with
respect to time is not given, but express the relative entropy in
the density contrast. The key step in this process is the validity
of Eq.(\ref{overdenseRE}). This step is believed to be correct.
The analysis above is applicable to the non-linear regime, too.
The results are summarized as follows. The equations
(\ref{densityConRE}) and
(\ref{derivativeOfFT1}),(\ref{derivativeOfFT2}) show that the
relative entropy, $S\{\rho||\langle\rho\rangle_D\}$, is an
increasing functional of the density contrast, $\delta$. And the
density contrast is an increasing function of time in linear
regime and the non-linear regime. So it is concluded that the
relative entropy is globally an increasing function of time in
both the linear regime and the non-linear regime.

Now a conjecture is given: The Relative Information Entropy of a
dust matter model $S\{\rho||\langle\rho\rangle_{\Sigma}\}$ is,
globally (i.e. averaged over the whole compact manifold $\Sigma$)
an increasing function of time.

Compared with the conjecture in \cite{g0402076}, in the conjecture
here, the condition, \emph{``for sufficiently large times"}, is
taken off.

This paper just give an illustration of the conjecture during the
large scale structure formation in cosmology. A strict proof need
be explored further.

\end{document}